\documentclass[12pt]{article}

 \usepackage{epsfig}
 \usepackage{amssymb}

 \def\M{\mathcal{M}} \def\MSQ{\overline{|\M|^2}}

 \def\prop{\ensuremath{\mathcal G}} \def\ckm{\ensuremath{V_{\rm CKM}^2}}
 \def\aem{\ensuremath{\alpha_{\rm EM}}} 
  
  \def\nc{\ensuremath{N_C}}
     
 \def\slash{\kern 1pt/\kern -6pt}
 \def\etmiss{\ensuremath{\kern 3pt\big{/}\kern -9pt E_T}}
  
  \def\ims #1 {\ensuremath{M^2_{[#1]}}}
 \def\sm{\ifmmode{{\rm SM}}\else{Standard Model}\fi}
 \def\qcd{\textsc{Qcd}}  
 \def\susy{\ifmmode{\rm SUSY}\else{supersymmetric}\fi}  
 \def\mssm{\ifmmode{\rm MSSM}\else{\textsc{Mssm}}\fi}
 \def\Susy{Supersymmetric} \def\MSSM{Minimal \Susy\ \sm}
  \def\vegas{\texttt{VEGAS}}

 \def\pl #1 #2 #3 {{\it Phys.\ Lett.} {\bf#1} (#2) #3}
 \def\np #1 #2 #3 {{\it Nucl.\ Phys.} {\bf#1} (#2) #3}
 \def\zp #1 #2 #3 {{\it Z.\ Phys.} {\bf#1} (#2) #3}
 \def\pr #1 #2 #3 {{\it Phys.\ Rev.} {\bf#1} (#2) #3}
 \def\prep #1 #2 #3 {{\it Phys.\ Rep.} {\bf#1} (#2) #3}
 \def\prl #1 #2 #3 {{\it Phys.\ Rev.\ Lett.} {\bf#1} (#2) #3}
 \def\mpl #1 #2 #3 {{\it Mod.\ Phys.\ Lett.} {\bf#1} (#2) #3}
 \def\rmp #1 #2 #3 {{\it Rev.\ Mod.\ Phys.} {\bf#1} (#2) #3}
 \def\jp #1 #2 #3 {{\it J.\ Phys.} {\bf#1} (#2) #3}     
 \def\cpc #1 #2 #3 {{\it Comp.\ Phys.\ Comm.} {\bf#1} (#2) #3}
 \def\epj #1 #2 #3 {{\it Eur.\ Phys.\ J.} {\bf#1} (#2) #3}
 \def\jhep #1 #2 #3 {{\it JHEP} {\bf #1} (#2) #3}
 \def\ibidem #1 #2 #3 {{\it ibidem} {\bf#1} (#2) #3}
 \def\xx #1 #2 #3 {{\bf#1}, (#2) #3}
 \def\preprint{{\it preprint}}
 \def\cavendish #1 {\preprint\ Cavendish--HEP--#1}

\begin{document}
\thispagestyle{empty}
\setcounter{page}{0}

\begin{flushright}
{RAL--TR--1999--018}\\
{February 1999\hspace*{.5 cm}}\\
\end{flushright}

\vspace*{\fill}

\begin{center}
{\Large \bf 
The phenomenology of top quark polarisation\\\vspace*{0.25cm}
in the decay of heavy charged Higgs bosons\\\vspace*{0.25cm}
at LHC and beyond}\\[1.5cm]
{\large Kosuke Odagiri}\\[0.4cm]
{\it Rutherford Appleton Laboratory,}\\
{\it Chilton, Didcot, Oxon OX11 0QX, UK.}\\[0.5cm]
\end{center}
\vspace*{\fill}

 \begin{abstract}
 {\noindent
 It has recently been shown that through the $\tau\nu$ decay mode the
heavy charged Higgs bosons $H^\pm$ can be discovered at LHC over a vast
region of the parameter space previously thought unaccessible.
 If the charged Higgs bosons are discovered in this decay mode and the
measurements of its mass and $\tan\beta$ are made, it allows us to make
greater use of the dominant but \qcd\ background contaminated $bt$ decay
mode by cuts suppressing phase space for the background.
 In this paper we study the top quark polarisation in the $bt$ decay mode. 
When $\tan\beta$ is either moderately large ($\tan\beta\gtrsim9$) or small
($\tan\beta\lesssim4$), the correlation effects can be used as a signal of
$H^\pm$ production.
 We outline the analysis procedure for various decay signatures. As an
example, we demonstrate the application of our procedure to the dominant
production process $gb\to tH^-$ at LHC.
 The effect is clearly observable for $H^\pm$ mass between 200 GeV and 500
GeV in both leptonic and hadronic channels, but at 500 GeV the boost on
the top quark starts to prevent the reconstruction of invariant masses,
and hence the observation of the correlation effect, in the hadronic decay
channel.
 }
 \end{abstract}

\vspace*{\fill}
\newpage

\section{Introduction}

 The charged Higgs bosons $H^\pm$ are a central ingredient of the Two
Higgs Doublet Model (2HDM) which is not only an attractive alternative to
the \sm\ single Higgs doublet but is also inherent in the supersymmetric
models, notably \MSSM\ (\mssm). Within the framework of \mssm, recent
experimental analyses \cite{lep2search, bdecay} point to the heavy
($M_{H^\pm}\gtrsim m_t$) mass region, where they can not be produced in
top quark decay, and their dominant decay $H^\pm\to bt$ suffers from large
\qcd\ background.

 In \cite{heavytaunu} the author has shown that the $\tau\nu$ decay mode
of the heavy charged Higgs bosons $H^\pm$ offers a striking signature of
high $p_T$ tau and large missing energy through which $H^\pm$ can be
discovered readily at LHC, provided that the top quark momentum in the
dominant production process $gb\to tH^-$ is constrained. Although the
signal cross section falls markedly at large $M_{H^\pm}$ and small
$\tan\beta$, it was claimed that for $\tan\beta\gtrsim3$ and
$M_{H^\pm}\lesssim1$ TeV the signal is sizeable enough after selection
cuts to compete with the irreducible background and the finite LHC
luminosity. 

 Given now that heavy $H^\pm$ can be discovered in the $\tau\nu$ decay
mode for much of the parameter space, and that $M_{H^\pm}$ and $\tan\beta$
can be measured to certain extent using the tau and missing transverse
momenta distributions and the total cross section, it is desirable to be
able to narrow down the parameters further by utilising the dominant decay
mode $bt$.

 The $bt$ decay mode of $H^\pm$, following production from the parton
level process $gb\to tH^-$, has been studied extensively in the literature
\cite{heavysearches} from the viewpoint of the search for heavy $H^\pm$.
The utility of the $bt$ decay mode is limited for this purpose because of
the contamination from \qcd\ background. However, given that $M_{H^\pm}$
and $\tan\beta$ can be measured in the $\tau\nu$ decay mode, this can be
used to constrain the final state.

 In this paper we study the effect of top quark polarisation from $H^\pm$
decay on parton level observables, noting that the top quark is polarised
by the chiral coupling of $H^\pm$ to $bt$, for either large or small
$\tan\beta$. To be more precise, the two chiral components of the $H^\pm$
coupling to $bt$ are proportional to $m_b\tan\beta$ and $m_t\cot\beta$ and
the two balance out when $\tan\beta=\sqrt{m_t/m_b}\sim 6.4$ ($m_b=4.25$
GeV). Far away from this central value, $\tan^4\beta\ll m^2_t/m^2_b$ or
$\tan^4\beta\gg m^2_t/m^2_b$, the couplings are dominantly left or right
leading to characteristic correlation patterns.

 Our discussions are applicable to any class of production processes, but
as an example for studying the effect of boosts on $H^\pm$ momenta we
adopt the $gb\to tH^-$ production process at LHC.

 When $\tan\beta\sim m_t/m_b$, the production cross section in the $gb\to
tH^-$ reaches the minimum, while the correlation pattern will resemble the
background, and our analysis will not be of much use. If the background
reduction is almost perfect before the polarisation analysis, the
polarisation measurement can be used to measure $\tan\beta$ accurately.
This is presumably unlikely.

 The use of right-handed tau polarisation as a characteristic signal of
$H^\pm$ decay was studied in \cite{bkm}. In addition, we note that if the
tau is produced in the decay of the top quark through $H^\pm$, the
distribution of invariant mass squared \ims{b\tau} \ is flat since there
is no momentum correlation between the $b$ from the decay of the top quark
and the $\tau$. This does not apply to the background from $W^\pm$ decay,
and if we know the exact distribution, we can use it to measure the signal
to background ratio and thus confirm the tau polarisation analysis.

 Our current work is in a similar vein. As in \cite{bkm}, the technique is
not expected to be of great use in reducing the background, but once an
excess of signal over background is observed near $H^\pm$ mass, the
momentum correlations resulting from top quark polarisations can be used
as a check for the signal presence in the sample and its proportion to the
background.

 We outline the analysis procedure for various decay signatures.


 Finally, we note that a preliminary study of the top quark polarisation
in $H^\pm$ decay was presented in \cite{czarpin}. Our present analysis is
different in several ways. Firstly, the primary purpose of \cite{czarpin}
is in measuring $\tan\beta$, rather than in distinguishing the signal from
the background. Thus their analysis is applicable in a range of
$\tan\beta$ complementary to the range emphasised here. Secondly, the
variable utilised in \cite{czarpin}, namely the leptonic energy in the
$H^\pm$ rest frame, is not well defined unless $H^\pm$ is produced at an
$e^+e^-$ collider and decays leptonically with no other source of missing
energy. In any case, for the intermediate region of $\tan\beta$ the
production rate at hadronic colliders is small.

\section{Calculation}

 We adopt $M_{H^\pm}$ and $\tan\beta$ as the two Higgs sector parameters.
 When calculating the width of $H^\pm$ in order to simulate the finite
width effect, we assume that the $bt$ branching ratio is 100\%. The top
branching ratio to $bW^\pm$ is also taken to be 100\%, and $W^\pm$
branching ratio to a pair of fermions $ff'$ is divided into quarks and
leptons by the tree level ratios of $1/3$ for each quark pair and $1/9$
for each lepton pair. Note that unlike in reference \cite{heavytaunu},
`lepton', $\ell$, always includes tau.

 We restrict our discussions to the tree level. The dominant part of the
radiative corrections \cite{oneloop}, which are known to be significant,
can be absorbed in terms of the running mass of the bottom quark and
therefore does not affect the shapes of our distributions. We simulate the
jet showering by imposing cuts between parton directions for the case of
the hadronic decay of the top quark.

 For simplicity we take the top quark mass to be 175 GeV for both Yukawa
coupling and the kinematics, and the bottom quark mass to be 4.25 GeV,
again for both Yukawa coupling and the kinematics. The fermions from
$W^\pm$ decay are considered massless.

 The electroweak parameters are $\aem=1/128$, $\sin^2\theta_W=0.2315$,
$M_Z=91.187$ GeV, $M_W=M_Z\cos\theta_W\approx79.94$ GeV.
 The Cabbibo--Kobayashi--Maskawa matrix element $\ckm[bt]$ is taken to be
1.

 The tree level matrix elements squared are simple and are calculated by
hand. Consistency checks are carried out between different limits. The
numerical integrations are carried out using \vegas\ \cite{vegas} and the
total decay width, in the narrow width limit of the propagators, agrees
with the width calculated by hand.

\section{Analysis}

 The decay matrix element squared, corresponding to the Feynman graph of
figure \ref{decayfey}, is given by:
 \begin{eqnarray}
 \MSQ&=&4\nc\left(\frac{e^2}{2\sin^2\theta_W}\right)^3
 |\prop_t\prop_{W^\pm}|^2(2p_b\cdot p_\nu)(2p_t\cdot p_\ell)
 \times\nonumber\\\label{fullmesq}&&\times
 \left[\left[g^2_b(M^2_{H^\pm}-m_b^2-p_t^2)-2g_bg_tm_bm_t\right]+
  \frac{2p_{\bar b}\cdot p_\ell}{2p_t\cdot p_\ell}(g_t^2m_t^2-g_b^2p_t^2)
 \right].
 \end{eqnarray}
 We have defined $g_t=(m_t\cot\beta/M_W)$ and $g_b=(m_b\tan\beta/M_W)$. 
$\prop_t$ and $\prop_{W^\pm}$ are Breit-Wigner propagators for the top
quark and $W^+$, respectively. In our analysis we also generate the
$H^\pm$ mass with a Breit-Wigner distribution.
 This formula essentially carries all the information about momentum
correlations. The effect of top quark polarisation is manifested in the
second term in large square brackets which is proportional to $(2p_{\bar
b}\cdot p_\ell)$. If the two chiral couplings were equal, the term
vanishes in the limit of on-shell top quark.

 In the limit of on-shell top quark and $W^+$ boson, the matrix element
squared, and hence the differential decay rate, depend only on two
independent dot products (or invariant masses), which we can take to be
$(2p_{\bar b}\cdot p_\ell)$ and $(2p_b\cdot p_\ell)$ since these two
quantities are measurable, as invariant masses, in the leptonic decay of
the top quark, if bottom quark charge tagging is available. We have:
 \begin{eqnarray}
 2p_b\cdot p_\nu  &=& m^2_t-m^2_b-M^2_{W^\pm}-2p_b\cdot p_\ell \nonumber\\
 2p_t\cdot p_\ell &=& M^2_{W^\pm}+2p_b\cdot p_\ell \label{dotproducts}
 \end{eqnarray}
 and hence the above statement holds. Furthermore, from equations
(\ref{fullmesq}) and (\ref{dotproducts}) we see that in the limit of
non-chiral coupling, the dependence on $(2p_{\bar b}\cdot p_\ell)$
vanishes altogether, as would a general non-chiral background. Hence the
dependence of the differential decay rate on $(2p_{\bar b}\cdot p_\ell)$
would be an indication of the presence of $H^\pm$. This would not hold if
the two coupling strengths were similar, $g_t^2\sim g_b^2$. This implies
$\tan^4\beta\sim m_t^2/m_b^2$.

 In experimental analyses, a possible approach would be to fit the
multi-dimensional differential decay distributions with theoretical
expectations for signal and background. However, for our purposes, let us
define the following dimensionless quantity in the limit of on-shell
$W^+$:
 \begin{equation}\label{lambda}
 \lambda 
 = \lambda_0\frac{\ims{\bar b\ell} -m_b^2}{\ims{b\ell} -m_b^2+M^2_{W^\pm}}
 = \lambda_0\frac{2p_{\bar b}\cdot p_\ell}{2p_b\cdot p_\ell+M^2_{W^\pm}}
 = \lambda_0\frac{2p_{\bar b}\cdot p_\ell}{2p_t\cdot p_\ell}.
 \end{equation}
 From (\ref{fullmesq}), the differential distribution of $\lambda$
translates directly to the chiral coupling strengths. $\lambda_0$ is the
normalisation which is a function of the masses. Given that $M_{H^\pm}$
needs to be known approximately for our analysis to be of much use, we can
define $\lambda_0$ for instance as follows:
 \begin{equation}\label{lambdanorm}
 \lambda_0 = \frac{m^2_t-M^2_{W^\pm}}{M^2_{H^\pm}-m^2_t} \approx
 \frac{(2p_b\cdot p_\ell)_{\rm max}}{(2p_{\bar b}\cdot p_\ell)_{\rm max}}.
 \end{equation}

 The dependence on $\lambda$ observed will be the sum of the signal and
the background which, after the signal selection cuts, is compatible with
$H^\pm$ production. If the background does not involve chiral
interactions, the distribution is flat, and even when the background does
involve chiral interactions, the distribution can be analysed as the
linear combination of the signal and background expectations. One can
compare the ratio thus obtained with the signal to background ratio
observed in order to confirm the presence of $H^\pm$ signal.

 In the next chapter we study the exact definitions of $\lambda$ for
different final states.

\section{Results}

\subsection{Leptonic decay with charge tagging}

 The simplest case is the leptonic decay of the top quark, with the charge
tagging of both bottom jets and the lepton. This allows us to assign all
momenta correctly. The size of the sample will be limited, since in
addition to the leptonic branching ratio of the top quark, the rate is
reduced further by the requirement of bottom quark charge tagging, the
efficiency of which one might roughly estimate to be around
$(10\%)^2=1\%$.

 In figure \ref{figa}a we show the differential distribution of $\lambda$
at $M_{H^\pm}=500$ GeV. As stated in the previous chapter, the
distribution is flat for $\tan\beta=\sqrt{m_t/m_b}\approx6.4$, and rapidly
approaches the large and small $\tan\beta$ limits away from this central
value. Let us concentrate on the two representative extreme values
$\tan\beta=1.5$ and 30, and the central value $\tan\beta=6.417$ from here
on.

 The finite energy resolution of detectors can lead to a finite spread in
the invariant mass distribution \cite{gunion}. The binning in our
distributions of $\Delta\lambda=0.1$ are fairly broad, corresponding to
roughly $0.1$ times the energy scale considered. For the quantity
$(2p_t\cdot p_\ell)$ in the denominator of (\ref{lambda}), this
corresponds to the energy spread of $0.1m_t\sim15$ GeV. This is similar to
the resolution obtainable at LHC \cite{lhc}.

 The dependence on $M_{H^\pm}$ is studied at $\tan\beta=1.5$ and
$\tan\beta=30$ in figure \ref{figa}b. We see that the scale normalisation
of (\ref{lambdanorm}) is satisfactory, and there is little residual
dependence on $M_{H^\pm}$.

\subsection{Leptonic decay without charge tagging}

 Once we drop the bottom quark charge tagging requirement, we need to
consider a new definition of $b$ and $\bar{b}$ in the procedure outlined
above. It is possible that the signal selection procedure has already
picked out one or the other of the two bottom jets as belonging to the
decay of the top quark. If this is the case, the assignment is again
trivial. From here on we assume that this is not the case.

 If $M_{H^\pm}\gg m_t$ the procedure is straightforward, as $\bar b$ and
the top quark decay products point back-to-back, even when we consider the
boost from the production process. In this case $2p_b\cdot p_\ell\lesssim
m_t^2-M_W^2$ and $2p_{\bar b}\cdot p_\ell=\mathcal{O}(M^2_{H^\pm})$ is
much greater than the other dot product. The author has verified that the
distributions look identical to figure \ref{figa}a if we simply take
$(2p_b\cdot p_\ell)$ to be the smaller of the two dot products.

 For general $M_{H^\pm}$, particularly when $(M^2_{H^\pm}-m_t^2)\sim
(m_t^2-M^2_{W^\pm})$ at $M_{H^\pm}\sim250$ GeV, the two dot products have
similar distributions. In this case the charge tagging discussed above is
an effective strategy. Here we proceed by summing over the two
distributions corresponding to the `right' and `wrong' assignments of
bottom quark jets.

 The resulting distributions are shown in figure \ref{figc} for
$M_{H^\pm}=200$ GeV. This is already sufficiently below the contaminated
region. Note that the normalisation is to two. Comparing with figure
\ref{figa}a we see that the region $\lambda<1$ is similar in two
distributions, but figure \ref{figc} has tails at $\lambda>1$ where most
of the wrong assignment lies. 

 This result indicates that the mis-assignment of bottom quarks, for
example from production processes, is unlikely to have significant effect
on our analysis, because of the different kinematic limits on the
invariant masses.

 We see that the flatness of the `background' $\tan\beta=\sqrt{m_t/m_b}$
case is retained here.

\subsection{Hadronic decay}

 If the top quark decays hadronically, the mass reconstruction allows one
to assign $b$ and $\bar b$ correctly to the two bottom quark jets. We
assume that this has been done already in the signal selection process. We
also assume that bottom quark jets are distinguishable from light quark
jets.
 However, the $\ell$ and $\nu$ of equation (\ref{fullmesq}) are now
indistinguishable unless we tag the charges of light quark jets as well as
the bottom quark jets.

 Here we choose to proceed by defining $\ell$ to be the jet which
minimises $p_b\cdot p_{\rm jet}$.

 The result is shown in figure \ref{figd}. The distribution is somewhat
flatter than the distributions in the leptonic channel, figures
\ref{figa} and \ref{figc}, reflecting the difficulty with which the light
quark jets can be assigned to the parent quarks. $M_{H^\pm}=500$ GeV is
adopted, but the distribution for different $M_{H^\pm}$ are found to be
similar.

 The problem of assigning the jets correctly can be seen in figure
\ref{fige} which compares the differential distributions in invariant
masses \ims{bq} \ and \ims{b\bar q'} . The distributions are roughly
independent of either $M_{H^\pm}$ or $\tan\beta$. We see that the two dot
products have similar profiles, and our procedure can not escape
contamination from the wrong assignment.

\subsection{The effect of cuts for the hadronic decay}

 The above discussions for the hadronic decay assume that the parton
directions are known from the jet distributions. In practice, the jets can
overlap when either top quark decay products being produced from
relatively light $H^\pm$ interfere with the $\bar b$ jet, or when the top
quark is boosted in heavy $H^\pm$ decay and the decay products are
clustered near each other. These cases lead to ambiguous definitions of
jet momenta.

 Let us study the effect of jet showering by introducing cuts on jet
separations.

 A typical azimuthal-pseudorapidity separation $\Delta R
=\sqrt{(\Delta\phi)^2+(\Delta\eta)^2}$ required to define jets and reduce
the inter-jet interference might be taken to be 0.7. In terms of the
azimuthal angle at zero rapidity this corresponds to 0.7 radians, or 40
degrees.

 Let us proceed by replacing $\Delta R$ by the angle between two jets in
the rest frame of $H^\pm$.

 We do not consider the effect of nonzero $H^\pm$ momenta. This is
sensitive to the selection cuts imposed in order to reduce the background.

 The ratio of events passing the cuts is shown in figure \ref{figf}. The
suppression is severe at large $M_{H^\pm}$ where the production cross
section also falls. From the events which do pass the separation cuts, it
is difficult to extract correlation effects. At $M_{H^\pm}=1$ TeV, for
example, the effects vanish altogether.

 The distribution after cuts is shown in figures \ref{figg} for
$M_{H^\pm}=200$ GeV and 500 GeV, to be compared with figure \ref{fige}
before cuts. The normalisations are to unity before cuts.

 The effect of cuts is found to be significant, but not prohibitive in
studying top quark polarisation.

\subsection{Production processes}

 In order to study the effect of boosts on $H^\pm$ momenta from the
production process, we consider the production process $gb\to tH^-$. The
distributions are expected to be modified slightly because of the boosts
on $H^\pm$ momenta acquired from the production. For the calculation of
the production process we proceed as in \cite{heavytaunu}.

 In figures \ref{figi} we show distributions analogous to \ref{figg} for
the hadronic decay after cuts. The cuts are now in terms of $\Delta R$ in
the collider frame rather than the angle between jets. The normalisation
is to unity after cuts for each distribution. We see that the
distributions are not significantly affected by the boost.

 The detector cuts on pseudorapidity, and any other separation cuts
involving jets arising from the production process, for example the top
quark in $gb\to tH^-$, can only affect the normalisation of the
distribution. The separation cuts between the $H^\pm$ decay products are
significant only because they directly affect the range of invariant
masses accessible for our analysis.

 For the sake of completeness, the author has run an analogous simulation
for the leptonic top quark decay with the cuts $\Delta R>0.7$ between the
bottom quark jets and $\Delta R>0.4$ between the lepton and the bottom
quark jets. The effect has been found to be small, as expected.

\section{Conclusion}

 In this paper we studied the effect on top quark polarisation of the
chiral charged Higgs boson $H^\pm$ coupling to fermions. We discussed the
optimum variables $\lambda$ which can be adopted in both leptonic and
hadronic decays of the top quark.

 The variable $\lambda$ we adopted assumes that $M_{H^\pm}$ is already
approximately known by the time we apply this analysis, either by the
observation of a peak in invariant mass of $t\bar b$, or by the analysis
of $H^\pm$ decay to $\tau\nu$ as mentioned in \cite{heavytaunu}.

 For the leptonic decay of the top quark, $\lambda$ can be defined such
that the distributions have the desired features of a clear momentum
correlation effect and flat distributions for the background. Charge
tagging the bottom quark jets and the lepton helps when $M_{H^\pm}\sim250$
GeV.

 The correlation effect is significant both for large and small
$M_{H^\pm}$.

 For the hadronic decay of the top quark, the momentum correlation effects
are less pronounced, but it is still strong enough to be used as a signal
for $H^\pm$ production.

 The distributions are affected significantly in the hadronic decay case
by the requirement of jet separations. This is especially the case for
heavy $H^\pm$.

 The boost on $H^\pm$ from the production process is found to give only a
small effect, in the case of $gb\to tH^-$ production mode. We expect our
analysis to be general to all production modes, at LHC and beyond.

 In conclusion, top quark polarisation is useful for confirming the
presence of $H^\pm$ in a sample of final states containing $t\bar b$ in
experimental analyses of near future, at $\tan\beta\gtrsim9$ and
$\tan\beta\lesssim4$, and up to $M_{H^\pm}\sim500$ in both hadronic and
leptonic channels.

\subsection*{Acknowledgements}

 I thank Stefano Moretti for discussions, for pointing out reference
\cite{czarpin} and for reading the manuscript. I also thank Kaoru Hagiwara
and Mike Seymour for advice and discussions.

\goodbreak

\clearpage\pagestyle{empty}

\subsection*{Figure captions}

 \begin{figure}[ht]
 \caption{The tree level Feynman graph for the decay $H^+\to t\bar b$ and
subsequent top quark and $W^+$ boson decays. The $H^-$ decay can be
obtained trivially from $H^+$ decay.} \label{decayfey}
 \caption{The differential distribution of $\lambda$ for the leptonic
decay of the top quark, with charge tagging. Normalisation is to unity. 
Lines were used instead of histograms for the sake of clarity. The binning
width is 0.1. (a -- top) $M_{H^\pm}=500$ GeV, for five different values of
$\tan\beta$. (b -- bottom) $\tan\beta=1.5$ and $\tan\beta=30$, for three
different values of $M_{H^\pm}$. The $\tan\beta=30$ distributions have
negative gradients.} \label{figa}
 \caption{The differential distribution of $\lambda$ for the leptonic
decay of the top quark, without charge tagging. $M_{H^\pm}=200$ GeV, for
three different values of $\tan\beta$. Normalisation is to two.}
\label{figc}
 \caption{The differential distribution of $\lambda$ for the hadronic
decay of the top quark, before cuts. $M_{H^\pm}=500$ GeV, for three
different values of $\tan\beta$.} \label{figd}
 \caption{The differential distribution of two invariant masses \ims{bq} \
and \ims{b\bar q'} , corresponding to `b--nu' and `b--lep' respectively in
the figure legends, in the hadronic decay of the top quark. 
$M_{H^\pm}=500$ GeV, $\tan\beta=1.5$.} \label{fige}
 \caption{The ratio of $H^\pm$ decay events which pass the jet separation
requirement, for three different values of $\tan\beta$ and
200 GeV $\le M_{H^\pm}\le$ 1 TeV.} \label{figf}
 \caption{The differential distribution of $\lambda$ for the hadronic
decay of the top quark, after cuts. (a -- top) $M_{H^\pm}=200$ GeV; (b --
bottom) $M_{H^\pm}=500$ GeV. The normalisation is to unity before cuts.}
\label{figg}
 \caption{The differential distribution of $\lambda$ for the hadronic
decay of the top quark, after cuts, convoluted with $g\bar b\to \bar tH^+$
production process. (a -- top) $M_{H^\pm}=200$ GeV; (b -- bottom)
$M_{H^\pm}=500$ GeV. The normalisation is to unity after cuts.}
\label{figi}
 \end{figure}

\clearpage

 \begin{figure}[p]
 \centerline{\epsfig{figure=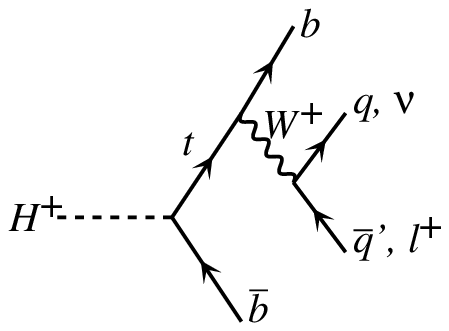,
           width=8cm,bbllx=0pt,bblly=0pt,bburx=130pt,bbury=120pt}}
 \centerline{Figure 1}
 \end{figure}

\clearpage

 \begin{figure}[p]
 \centerline{\epsfig{figure=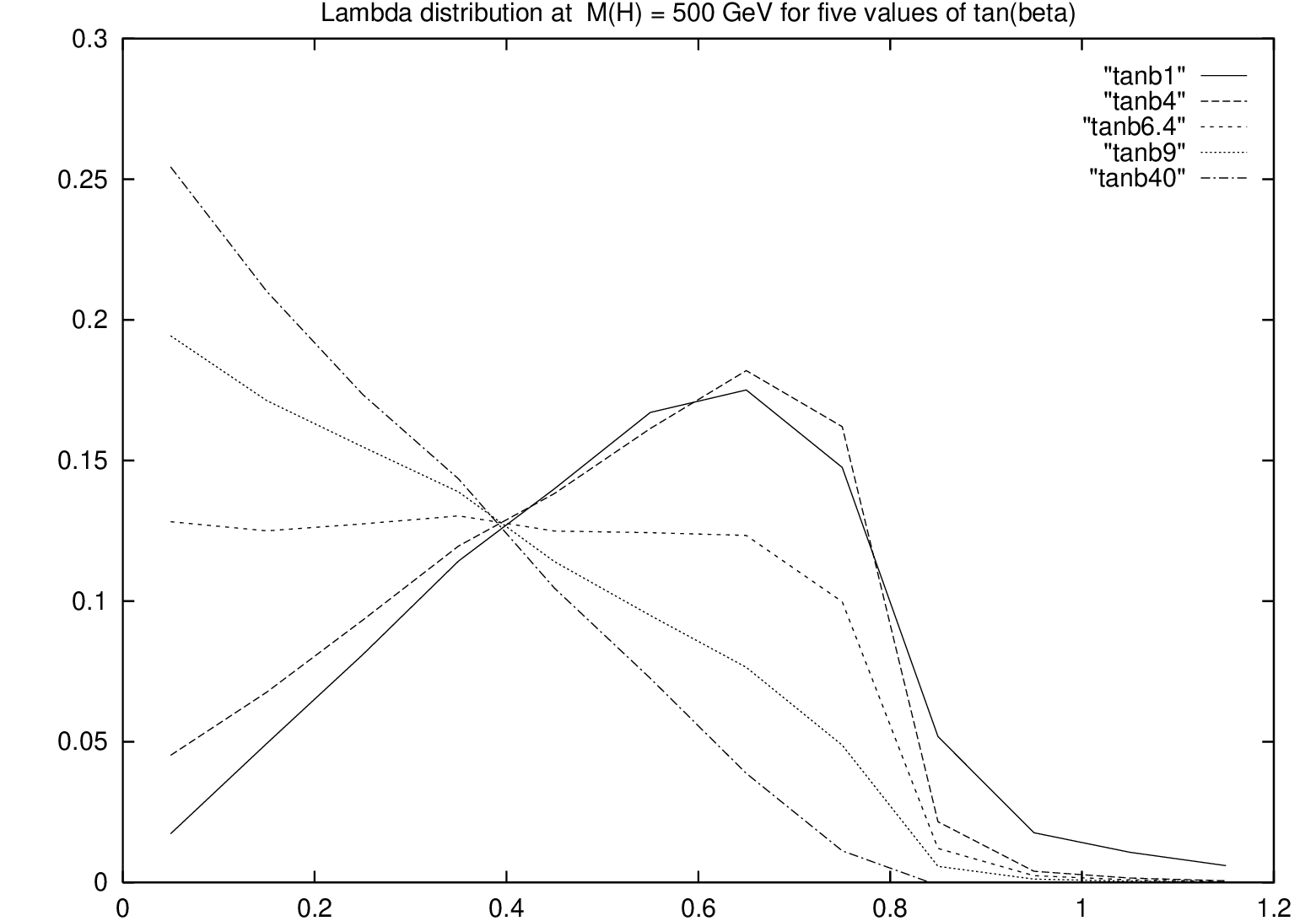, width=14cm}}
 \vspace*{0.5cm}
 \centerline{\epsfig{figure=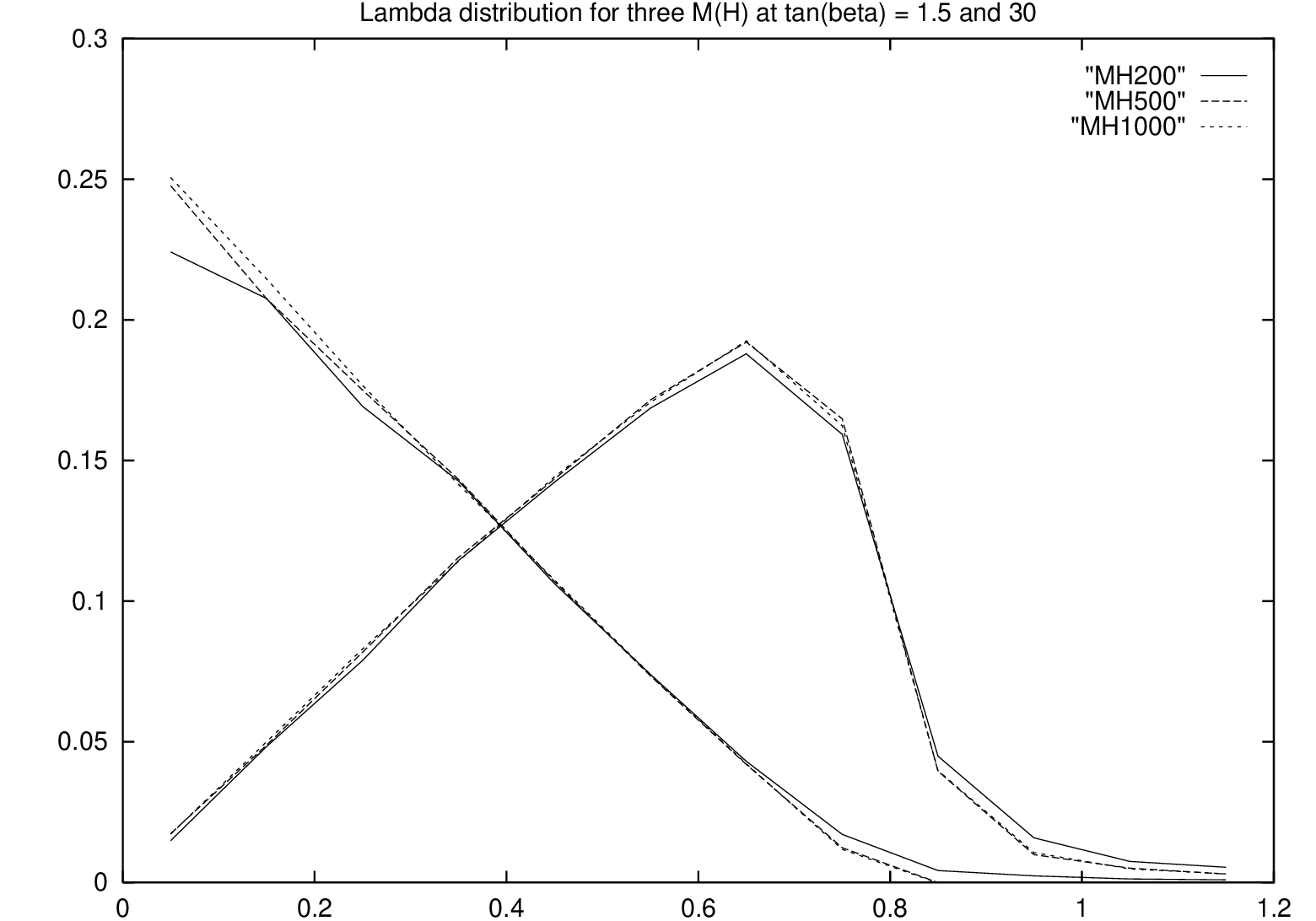, width=14cm}}
 \vspace*{0.5cm}
 \centerline{Figure 2}
 \end{figure}

\clearpage

 \begin{figure}[p]
 \centerline{\epsfig{figure=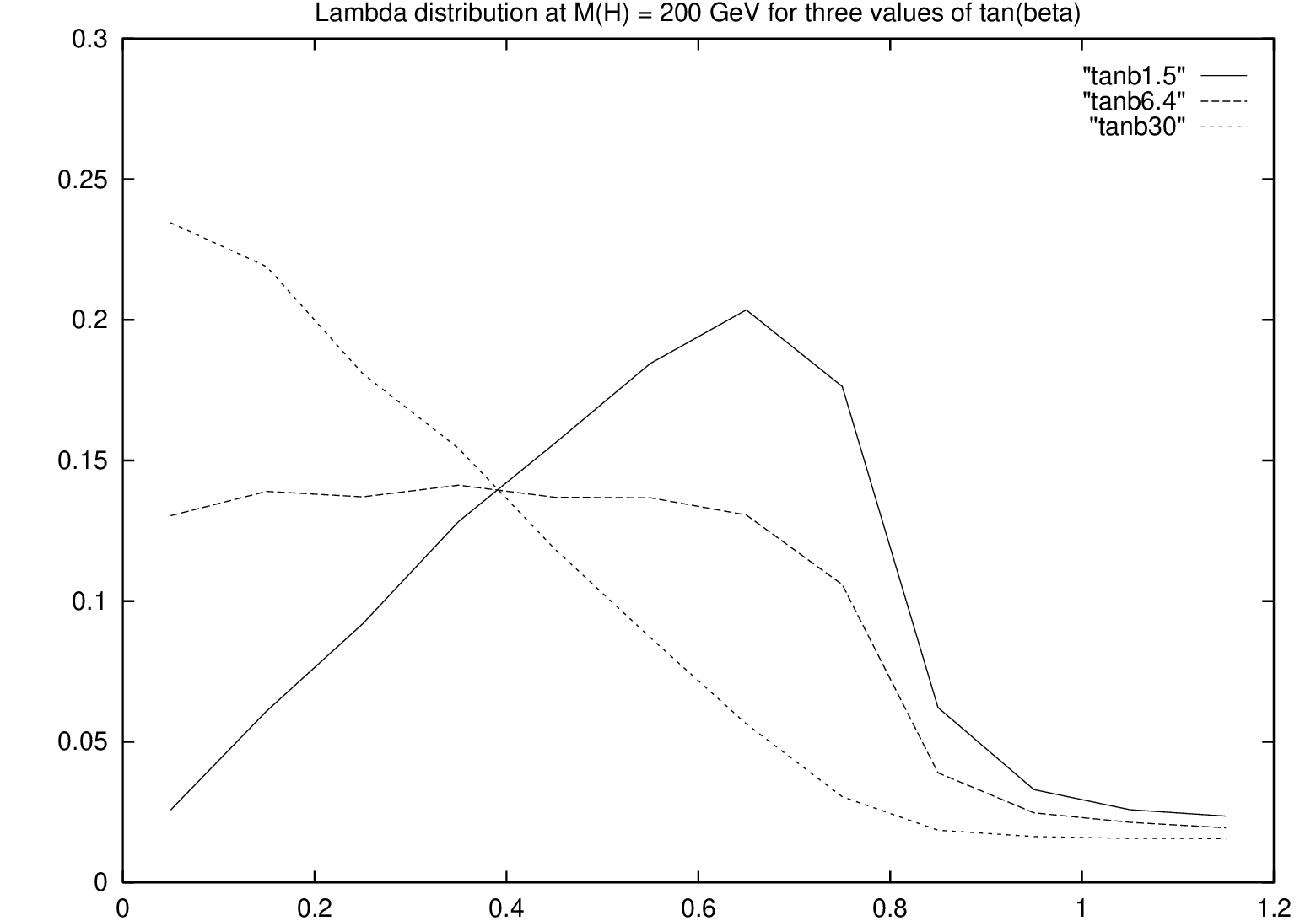, width=14cm}}
 \vspace*{0.5cm}
 \centerline{Figure 3}
 \end{figure}

\clearpage

 \begin{figure}[p]
 \centerline{\epsfig{figure=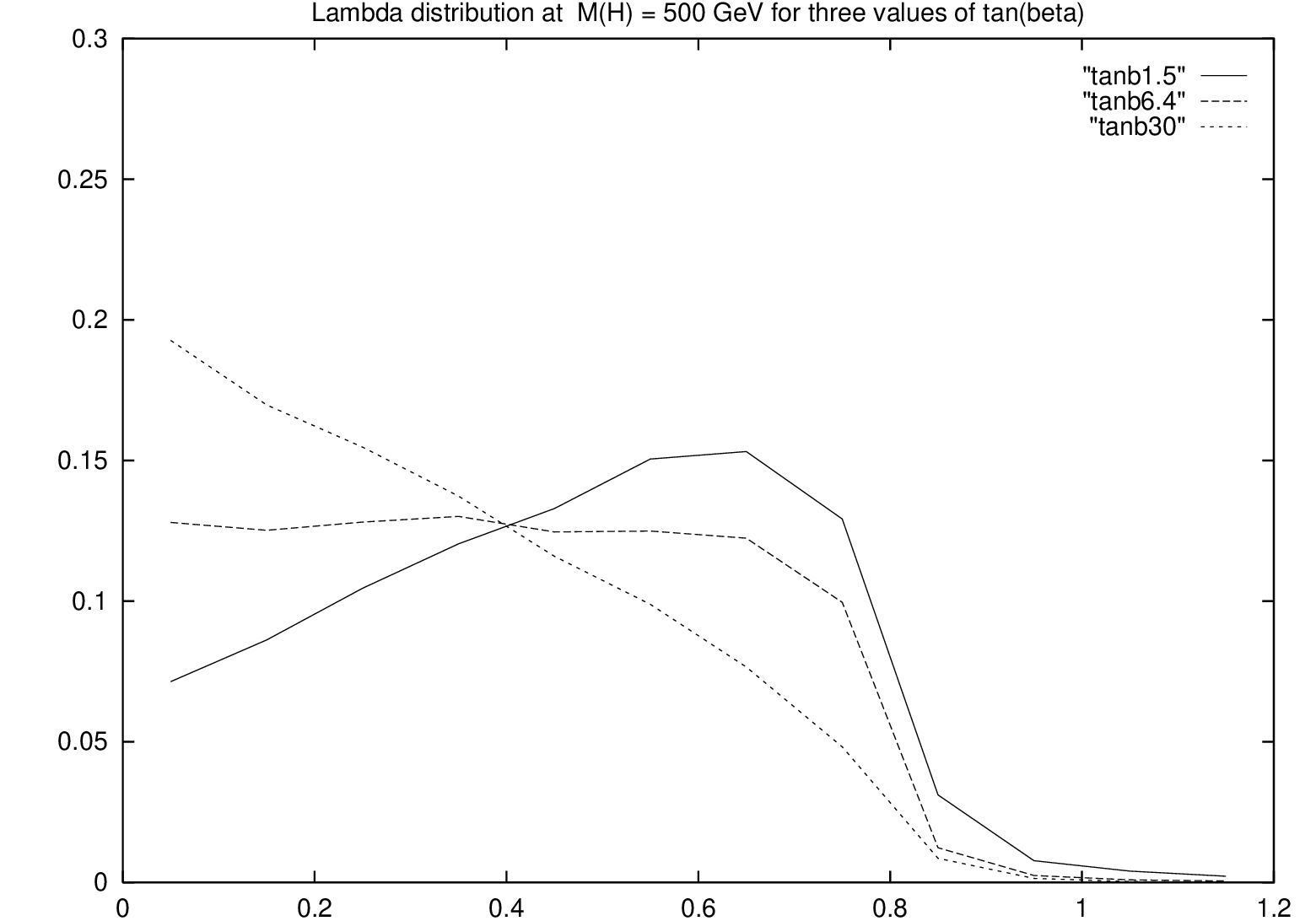, width=14cm}}
 \vspace*{0.5cm}
 \centerline{Figure 4}
 \end{figure}

\clearpage

 \begin{figure}[p]
 \centerline{\epsfig{figure=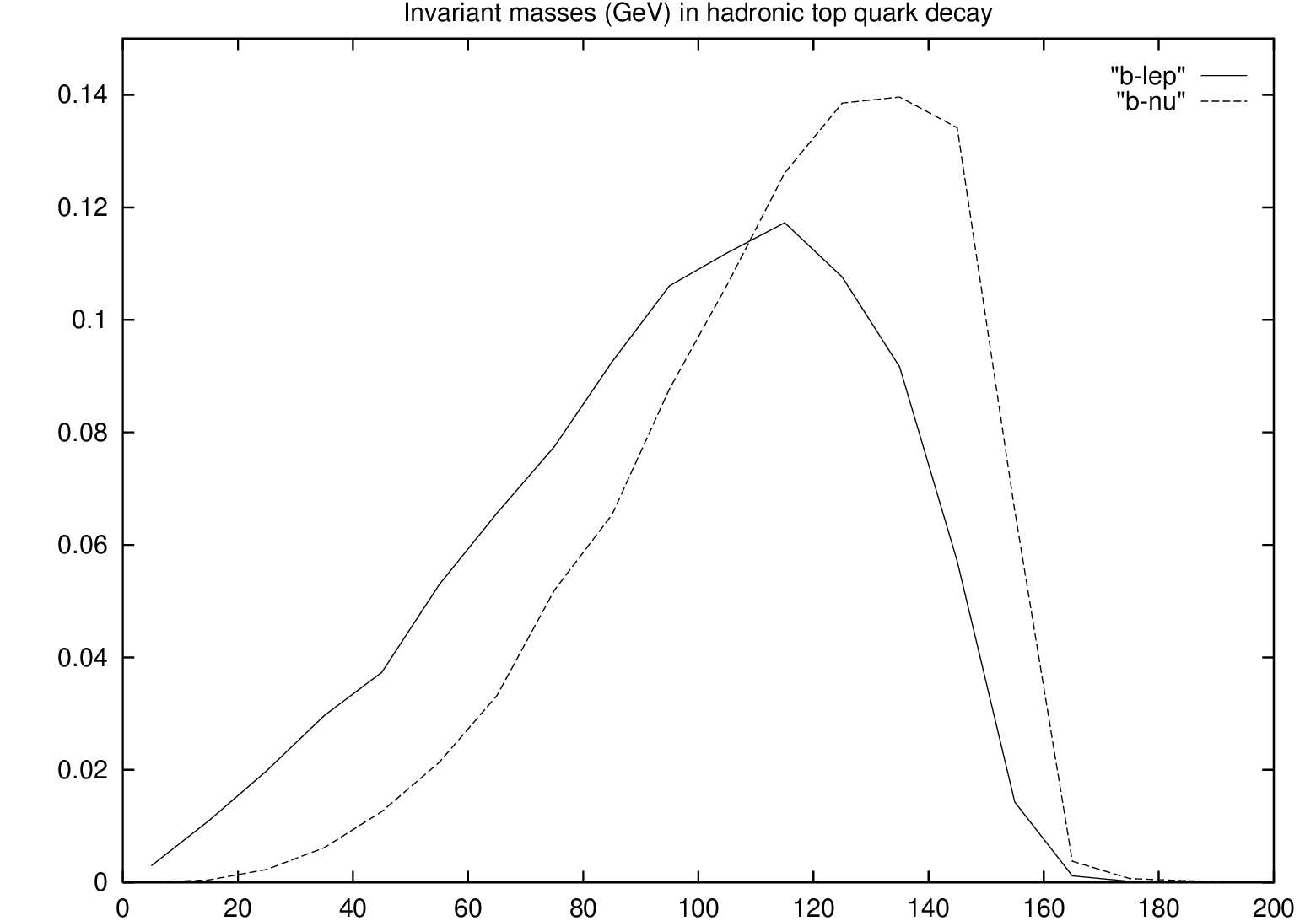, width=14cm}}
 \vspace*{0.5cm}
 \centerline{Figure 5}
 \end{figure}

\clearpage

 \begin{figure}[p]
 \centerline{\epsfig{figure=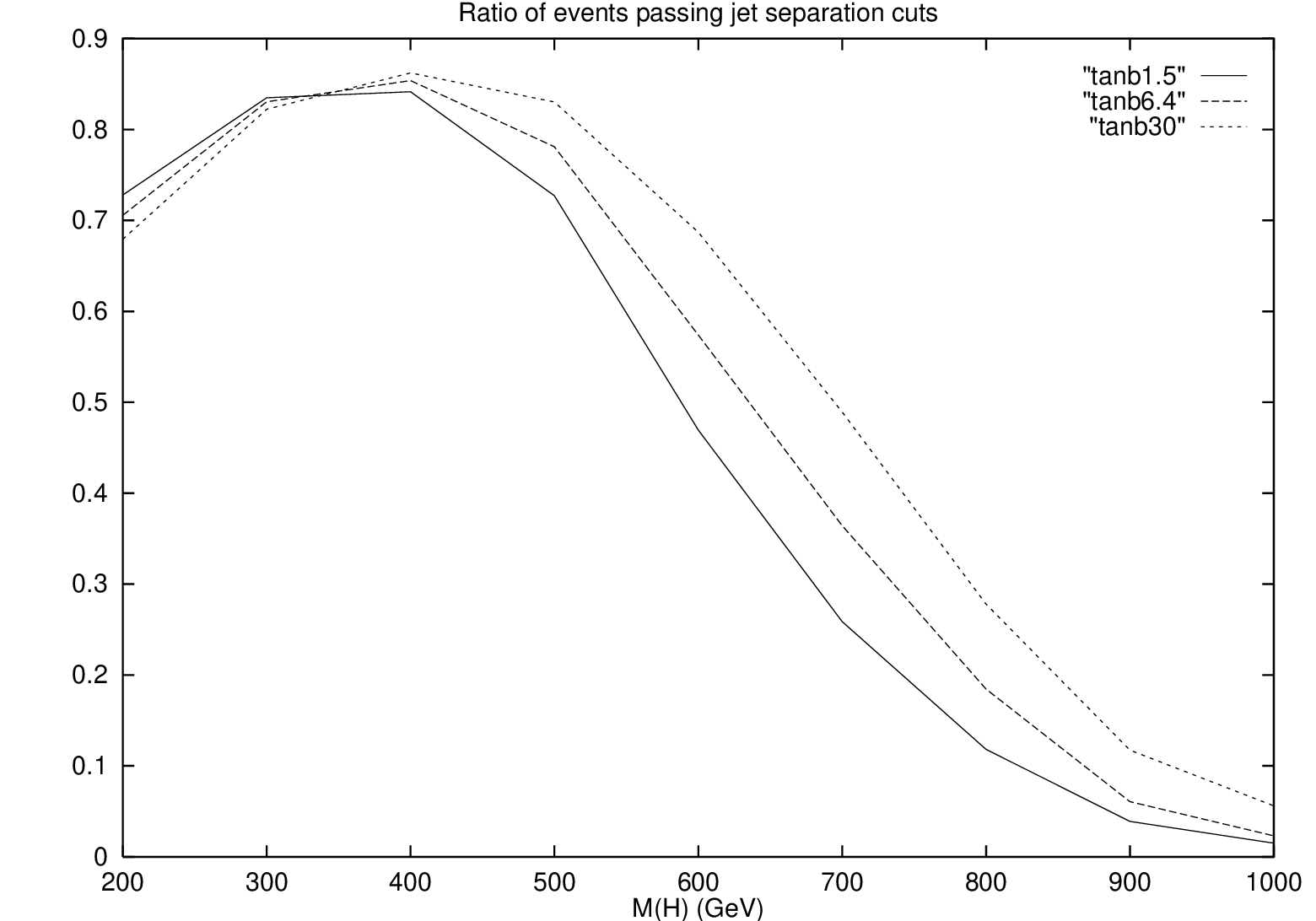, width=14cm}}
 \vspace*{0.5cm}
 \centerline{Figure 6}
 \end{figure}

\clearpage

 \begin{figure}[p]
 \centerline{\epsfig{figure=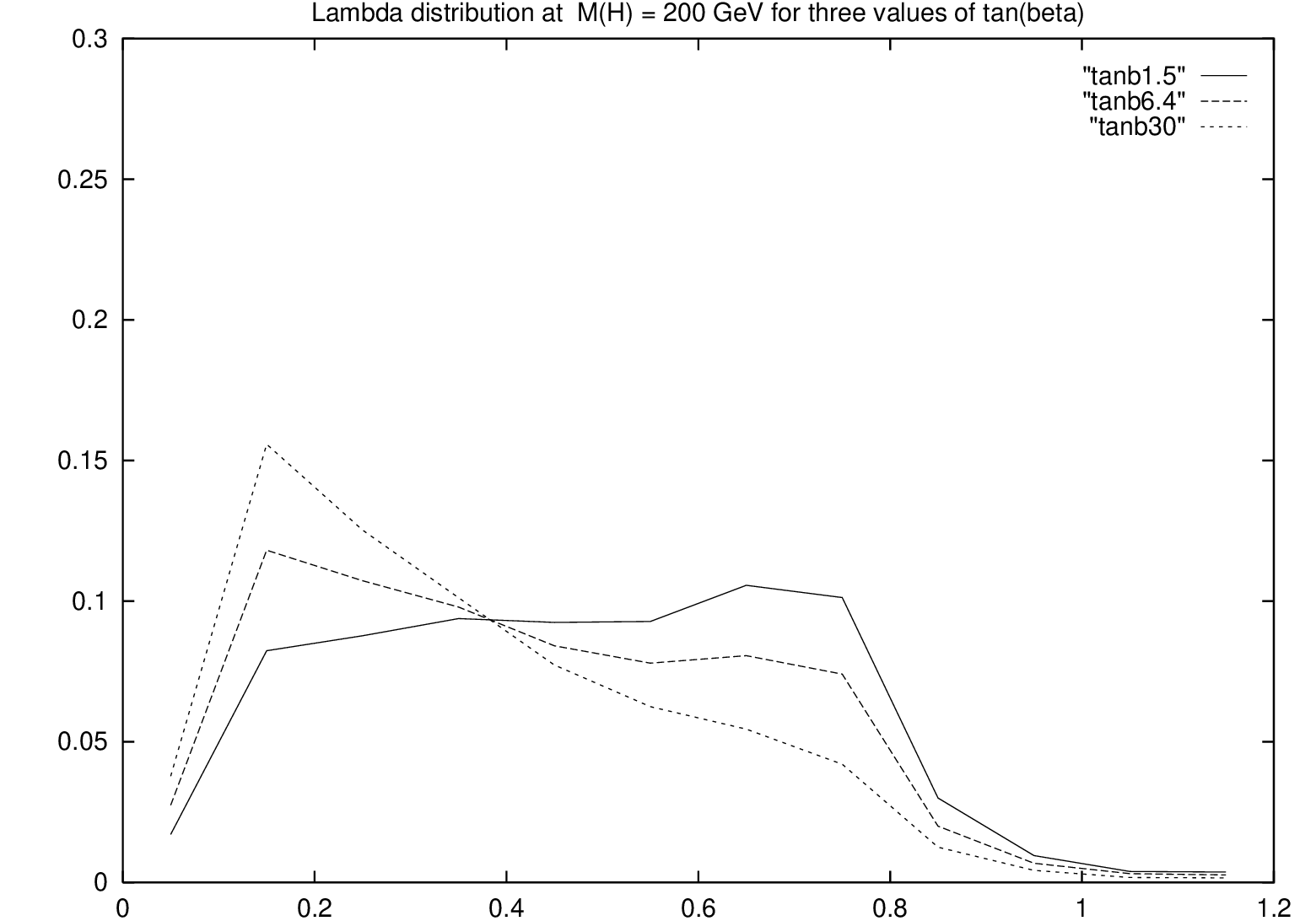, width=14cm}}
 \vspace*{0.5cm}
 \centerline{\epsfig{figure=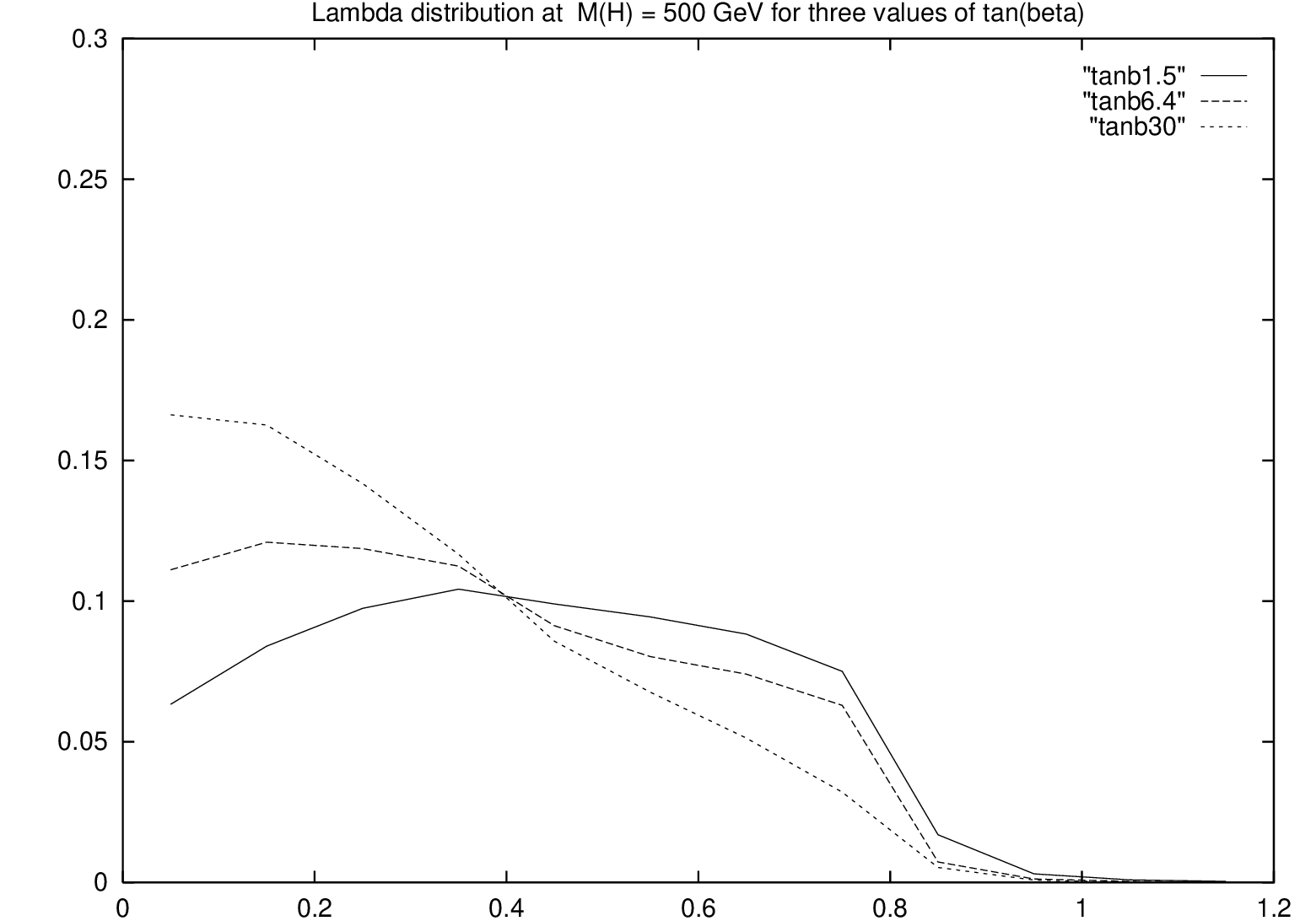, width=14cm}}
 \vspace*{0.5cm}
 \centerline{Figure 7}
 \end{figure}

\clearpage

 \begin{figure}[p]
 \centerline{\epsfig{figure=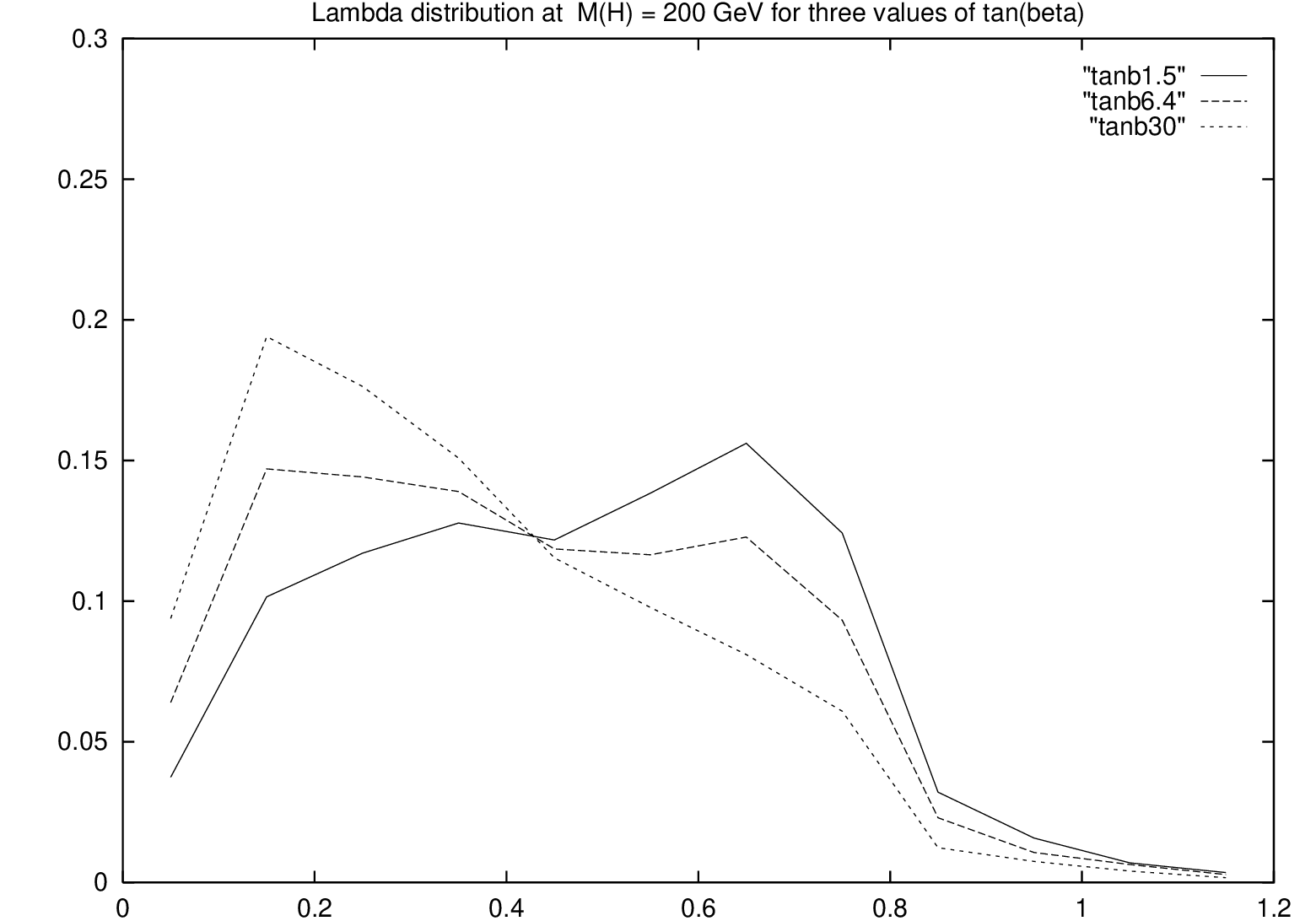, width=14cm}}
 \vspace*{0.5cm}
 \centerline{\epsfig{figure=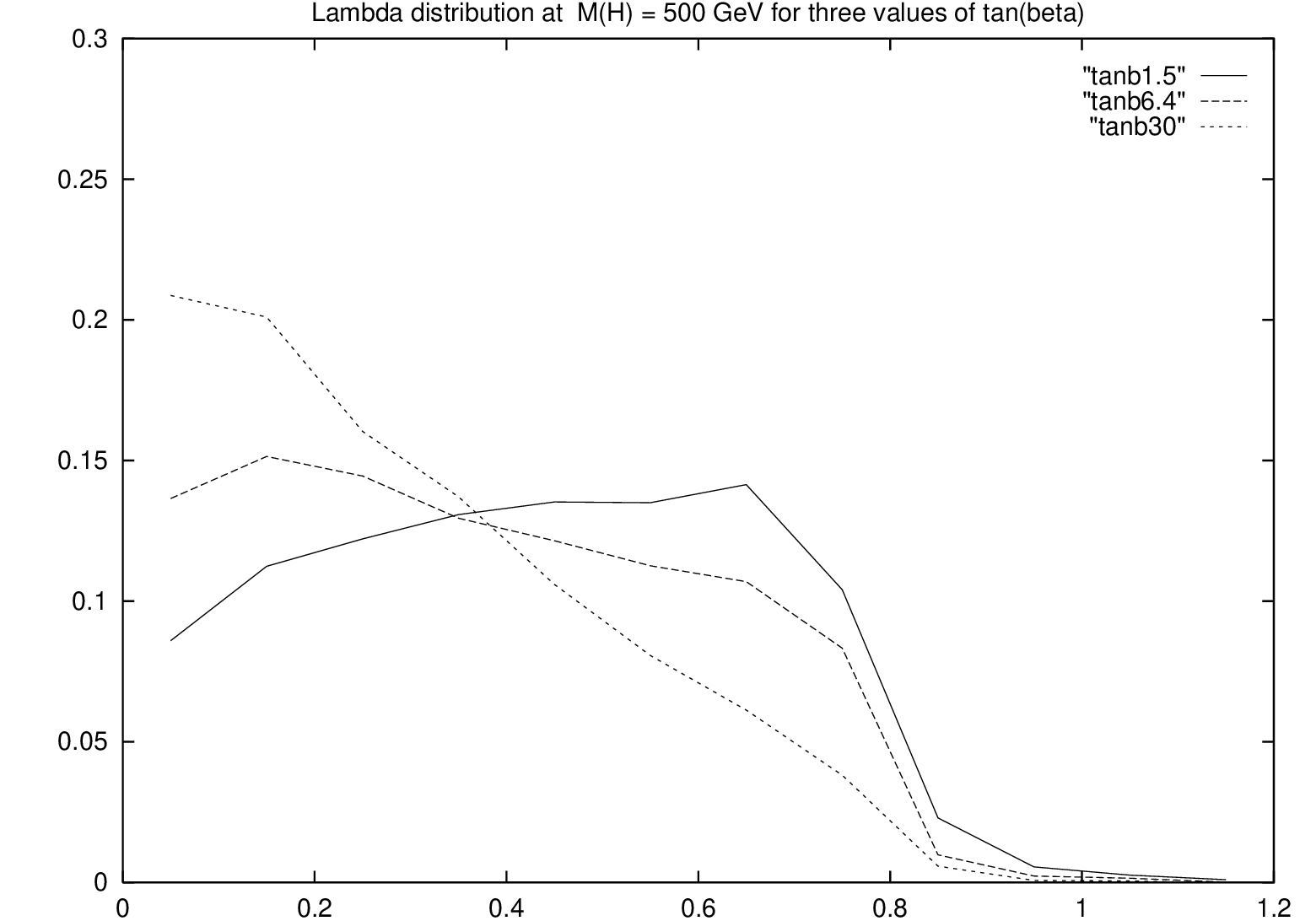, width=14cm}}
 \vspace*{0.5cm}
 \centerline{Figure 8}
 \end{figure}


\begin{thebibliography}{99}

\bibitem{lep2search} F.~Richard, \preprint\ LAL 98--74, hep-ex/9810045,
talk given at Zuoz Summer School on Hidden Symmetries and Higgs Phenomena,
Zuoz, Switzerland, 16--22 August 1998.

\bibitem{bdecay} The ALEPH collaboration, \pl B429 1998 169.

\bibitem{heavytaunu} K.~Odagiri, \preprint\ RAL--TR--1999--012,
hep-ph/9901432.


\bibitem{heavysearches}
 A.~Krause, T.~Plehn, M.~Spira and P.M.~Zerwas, \np B519 1988 85; \\
 V.~Barger, R.J.N.~Phillips and D.P.~Roy, \pl B324 1994 236; \\
 M.~Guchait and D.P.~Roy, \pr D55 1997 7263.

\bibitem{bkm} B.K.~Bullock, K.~Hagiwara and A.D.~Martin, \prl 67 1991
3055.

\bibitem{czarpin} A.~Czarnecki and J.L.~Pinfold, \pl B328 1994 427.

\bibitem{oneloop} A.~Mendez and A.~Pomarol, \pl B252 1990 461, \ \ibidem
B265 1991 177; \\ A.~Djouadi and P.~Gambino, \pr D51 1995 218.

\bibitem{vegas} G.P.~Lepage, {\it Jour.~Comp.~Phys.} \xx 27 1978 192.

\bibitem{gunion} J.F.~Gunion, \pl B322 1994 125.

\bibitem{lhc}
 CMS Technical Proposal, CERN/LHC/94-43 LHCC/P1 (December 1994); \\
 ATLAS Technical Proposal, CERN/LHC/94-43 LHCC/P2 (December 1994).

\end{thebibliography}
\end{document}